\documentclass[global,twocolumn]{svjour}
\usepackage{graphics}
\usepackage{epsfig}

\usepackage{amsmath}
\usepackage{amsopn}
\usepackage{amsfonts}
\usepackage{amssymb}
\journalname{Appl.Phys.B}

\newcommand{\ket}[1]{\left|#1\right>}

\begin{document}
%
%
%
%
\title{How to realize a universal quantum gate with trapped ions}
\author{Ferdinand~Schmidt-Kaler \thanks{corresponding author}, Hartmut H\"affner,
Stephan Gulde, Mark Riebe,  Gavin P.T. Lancaster, Thomas Deuschle,
Christoph Becher, Wolfgang H\"ansel,  J\"urgen Eschner, Christian
F. Roos, and Rainer Blatt }
 \authorrunning{F.~Schmidt-Kaler et al.}
 \institute{Institut f\"ur
Experimentalphysik, University of Innsbruck, A-6020 Innsbruck,
Austria
}                     
%
%
\institute{Institut f\"ur Experimentalphysik, Universit\"at
Innsbruck,
Technikerstra{\ss}e 25, A-6020 Innsbruck, Austria 
}
\email{ferdinand.schmidt-kaler@uibk.ac.at,  FAX: +435125072952}
\date{Received: date / Revised version: date}
%
\keywords{Quantum computing, quantum bits, entanglement, single
ions}

 \maketitle
%
\begin{abstract}
We report the realization of an elementary quantum processor based
on a linear crystal of trapped ions. Each ion serves as a quantum
bit (qubit) to store the quantum information in long lived
electronic states. We present the realization of single-qubit and
of universal two-qubit logic gates. The two-qubit operation relies
on the coupling of the ions through their collective quantized
motion. A detailed description of the setup and the methods is
included.
\end{abstract}
%

%
%

%
%
%
%
\section{Introduction}

Quantum computers (QC) are known to perform certain computational
tasks more efficiently than their classical counterparts. The
theoretical concept of QC is highly developed. Most well-known
among the quantum algorithms\cite{CHUANG00} is the efficient
algorithm for the factorization of large numbers\cite{SHOR97}
which threatens the security of the commonly used RSA-encryption
scheme. Furthermore, efficient quantum algorithms exist for
searching entries in an unsorted data base\cite{GROVER97}, for
simulating quantum spin systems\cite{JANE03}, and for quantum
games. As in a classical computer, errors will necessarily occur.
Although the nature of errors is different in quantum mechanical
and in classical computers, algorithms have been developed which
can correct qubit errors\cite{SHOR95,STEANE96}. World-wide efforts
aim at a scalable realization of a QC\cite{ARDA}. Already in 1995,
J.~I.~Cirac and P.~Zoller proposed to implement a scalable QC on a
string of trapped ions, where each ion's electronic state
represents a qubit\cite{CIRAC95}. Quantum gates between any subset
of ions would be induced by laser-ion interactions, including the
coupling of the ions to their collective quantized
motion\cite{SASURA02}. Today, a number of different proposals for
quantum gates in an ion based QC are known.

While the construction of a large scale QC might still be in
remote future, we may already today perform experiments with a
small number of qubits, bringing into reality what used to be
Gedanken experiments and thus enlightening the foundations of
quantum mechanics. This will serve to further extend our knowledge
of the puzzling quantum theory and its borderline to classical
physics, given by decoherence and the measurement
process\cite{DECOHERENCE98}.

The ion-trap system itself is fully understood theoretically, and
equally well its interaction with a laser field. Any kind of
quantum logic gate operation may thus be predicted. Actual
experiments are performed with few ions that are confined in a
Paul trap, such that time scales for decoherence and for the
dephasing of qubits due to fluctuations of external parameters are
long as compared to the coherent qubit operation times. The
detection of the ions' internal states relies on electron
shelving, leading to a detection efficiency near unity. In this
kind of fully defined, text-book like setting, elementary quantum
processors may be realized. Quantum logic gate operations and
entangled states may be studied.

The most challenging experimental step towards achieving the
Cirac$\&$Zoller scheme  (CZ) of a QC is to implement the
controlled-NOT (CNOT) gate operation between two individual ions.
The CNOT quantum logical gate corresponds to the XOR gate
operation of classical logic which flips the state of a target bit
conditioned on the state of a control bit. Taking the basis states
$|a,b\rangle$ = \{$|0,0\rangle, |0,1\rangle, |1,0\rangle,
|1,1\rangle$\} of two qubits, the CNOT operation reads
$|a,b\rangle \rightarrow |a,a \oplus b\rangle$, where $\oplus$
represents an addition modulo 2. Only if the control qubit (first
entry) is in $|1\rangle$, the quantum state of the control qubit
changes. Here, we present the realization of a CNOT quantum gate
\cite{SCHMIDTKALER03} according to the original CZ proposal
\cite{CIRAC95}.

In our experiment, two $^{40}$Ca$^+$ ions are held in a linear
Paul trap and are individually addressed with focussed laser
beams. Superpositions of long-lived electronic states represent a
qubit. By initializing the control and target qubit in all four
basis states and performing the CNOT operation, we determine the
desired truth table. To prove the quantum nature of the gate, we
use a superposition state for the control qubit and generate an
entangled output state.

The paper gives a detailed description of the experimental
apparatus and the required procedures in sections \ref{EXP_SETUP}
and \ref{PROCEDURES}. In sect.~\ref{TWOIONGATE}, we discuss the
realization of the universal two-ion CNOT gate, followed by a
discussion of its current limitations and possible future
improvements.

\section{Experimental setup \label{EXP_SETUP}}
\subsection{Levels and transitions in the $^{40}$Ca$^+$ ion}

\begin{figure}[t]
\begin{center}
\epsfig{file=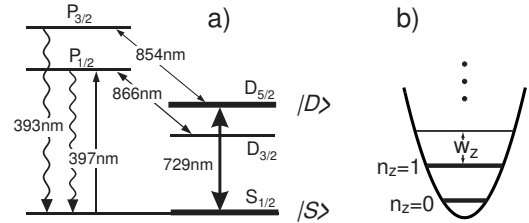,width=.8\linewidth} \caption{
a)$^{40}$Ca$^+$ level scheme. A qubit is encoded in the $S_{1/2},
(m=-1/2)$ ground and $D_{5/2}, (m=-1/2)$ metastable state of a
single trapped ion. b) The lowest two number states $n$ of an
axial vibrational motion in the trap are used as quantum bus.
\label{levels_f1} }
\end{center}
\end{figure}

The Calcium ion ($^{40}$Ca$^+$) has a single valence electron and
no hyperfine structure, see fig.~\ref{levels_f1}a for the relevant
levels and transitions. We have chosen $^{40}$Ca$^+$ for several
reasons: (a) The transition wavelengths for Doppler-cooling and
optical pumping are well suited for solid-state and diode laser
sources. (b) Long-lived metastable states ($\tau\sim1$\,s) allow
for the implementation of qubits. (c) The narrow-line quadrupole
transition can also be used to implement sideband cooling to the
vibrational ground state.

We cool the ion on the S$_{1/2}$ to P$_{1/2}$ transition near
397~nm close to the Doppler limit. The UV-radiation is produced as
the second harmonic of a Ti:Sapphire laser at 794\,nm\footnote{The
practicability of a grating stabilized
UV-diode\cite{HAYAS00,LANCESTER03} for single ion cooling and
detection has been proven.}. Grating stabilized diode lasers at
866\,nm and 854\,nm prevent pumping into the D$_{3/2}$ and
D$_{5/2}$  states. Each of the above lasers is frequency-locked to
its individual optical reference cavity using the
Pound-Drever-Hall method \cite{DREVER83}. With cavity linewidths
of 2-5\,MHz, we reach a laser frequency stability of better than
300\,kHz. Frequency tuning of the  lasers is achieved by scanning
the length of the corresponding reference cavities using
piezo-electric actuators.

The electronic level S$_{1/2}(m=-1/2) \equiv |S\rangle$ is
identified with logic $|0\rangle$ and D$_{5/2}(m=-1/2) \equiv
|D\rangle$ with logic $|1\rangle$, respectively. To perform
quantum logic operations, we excite the corresponding transition
with a Ti:Sapphire laser near 729\,nm. The complete laser system
for the qubit manipulation is described in
sect.~\ref{LaserSetup729} and \ref{AddressingOptics}.

We detect the quantum state of the qubit by applying the laser
beams at 397\,nm and 866\,nm and monitoring the fluorescence of
the ion at 397\,nm on a photomultiplier and on a CCD camera
(electron shelving technique\cite{DEHMELT75}). The internal state
of the ion is discriminated with an efficiency close to 100$\%$,
details of the detection are found in sect.~\ref{QubitReadout}.

It is of advantage that pure $^{40}$Ca$^+$ ion crystals can be
loaded into the trap using a relatively simple photo-ionization
scheme\cite{KJAERGARD00} that relies on a two step laser
excitation: A weak beam of neutral Ca is emitted by a resistantly
heated oven\cite{ROTTER}. Calcium atoms are excited on the
$4s^1S_0 \rightarrow 4p^1P_1$ transition near 423\,nm by a grating
stabilized diode laser\cite{GULDE01,ROTTER}. Ionization is reached
with radiation at $\lambda \leq$ 390\,nm using a UV-diode laser or
even a simple UV-light emitting diode.

\begin{figure}[t]
\begin{center}
\epsfig{file=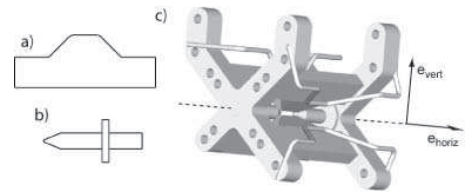, width=0.7\linewidth} \caption{Construction
of the linear trap\cite{GULDE_DR} out of four blades (a) and two
tips (b). The 3D-view (c) shows the arrangement of the RF-blades
which generate the radial trapping potential. The closest distance
between the  blades is 1.6\,mm. The tips are separated by 5.0\,mm.
All electrodes are mounted onto a Macor ceramics spacer. The
typical machining precision of all parts is 5 to 10\,$\mu$m. The
RF-blades are fabricated by electro-erosion from stainless steel,
the tips are made of molybdenum.}\label{Construction}
\end{center}
\end{figure}

\subsection{Linear Paul trap \label{PaulTrap}}

For the experiments, $^{40}$Ca$^+$ ions are stored in the harmonic potential
of a linear Paul trap. The trap is made of four blades for radial confinement
and two tips for axial confinement, see fig.~2. Under typical operating
conditions we observe axial and radial motional frequencies
$(\omega_\text{ax}, \omega_\text{rad})/2\pi=$~(1.2, 5.0)\,MHz, respectively.
The trap combines good optical access with relatively high trapping
frequencies, even though the trap dimensions are comparatively large.
Electrically insulating parts have no direct line of sight to the ions. We
attribute the low heating rate ($<$1\,phonon/50ms)\cite{ROHDE01} to the
combination of a large distance between ions and trap electrodes
(r$_0$$\approx$\,0.8\,mm) and the clean loading scheme by photo-ionization.
Both tips, typically at +1kV, are positioned in the symmetry axis with high
precision. Small asymmetries are compensated by applying voltages of below
200\,V to electrodes which are placed at a radial distance of 30\,mm from the
trap symmetry axis (fig.~\ref{Construction}c). The radio frequency (RF)
$\Omega/(2\pi)\simeq$ 23.5\,MHz is applied to two diagonally opposing blades
(the other two being at 0\,V). This creates an oscillating electrical
quadrupole field which results in a radial trapping potential. The RF is
generated by a synthesizer\footnote{Marconi Inc., Signal gen. 2019A} and
amplified\footnote{Minicircuits Inc., LZY-1} to 15\,W. A helical
$\lambda/4$-resonator (loaded Q-value $\sim$200) serves to match the
capacitive load of the trap structure with the 50\,$\rm \Omega$ output of the
amplifier and to enhance the drive voltage to a few kV$_{pp}$. We typically
operate the trap close to the stability parameter q$\leq$0.6\cite{GOSH95}. In
order to avoid RF pick-up on the DC-voltage leads we use separate
feed-throughs and filter the DC voltages.

The trap is mounted in a UHV housing, pumped by a Titanium
sublimation and an ion getter pump\footnote{Varian Inc.,
Starcell~20}. The residual gas pressure is below $2\times
10^{-11}$\,mbar.

\subsection{Optical setup \label{OpticalSetup}}

The output of a Ti:Sapphire laser\footnote{Coherent Inc., 899-21}
near 794~nm is frequency-doubled\footnote{Spectra Inc., Wavetrain}
to obtain up to 50~mW light at 397~nm. We stabilize the UV power
to 1$\%(rms)$ using an AOM in front of the doubling cavity. During
a gate operation on the qubit transition, any residual UV-light
has to be suppressed to a maximum. As the UV-light needs to be
switched faster than mechanical shutters would allow, we pass it
through an AOM\footnote{Brimrose Inc., QZF-80-20}, couple into a
single-mode polarization-maintaining fibre\footnote{Sch\"after
Kirchhof Inc.} and transport it to the trap.
After the fiber output, the light is sent through a second AOM.
Switching the RF-drive of both AOM's yields an extinction of about
$2\times10^{-6}$. Additionally, due to the fibre, the UV-beam is
spatially filtered such that its focus on the ion crystal
($w_0\sim50\mu$m, $\leq$100\,$\mu$W) does not cause excessive
stray-light on the trap electrodes. The UV-beam leaving the second
AOM is split into two beams which are superimposed with light at
866~nm and 854~nm. These beams enter the vacuum system via UV-AR
coated windows, and intersect at the ion trap under angles of
$\{$-22$_{hor.}^\circ$, 0$_{vert.}^\circ\}$ with respect to the
axial trap direction, and $\{$22$_{hor.}^\circ$,
45$_{vert.}^\circ\}$, respectively. The combination of both light
fields is used for Doppler cooling, ion detection, and the
compensation of micro-motion.

Another part of the UV-light transmitted through the fiber is
controlled by a third AOM, enters along the axis of the magnetic
field $\{$22$_{hor.}^\circ$, 0$_{vert.}^\circ\}$, and is applied
for optical pumping.

The switching of the light field at 854~nm is controlled by an
additional AOM in double-pass configuration to assure
on/off-dynamics of about $2\times10^{-4}$. The laser field at
866\,nm does not couple to the qubit levels and is kept on
continuously.

The fluorescence of the ions at 397~nm is collected through a
viewport using a large collimating
lens\footnote{Nikon,~MNH-23150-ED-Plan-1.5x} at a working distance
of 65~mm and focused onto an intensified CCD
camera\footnote{Princton Instum., Inc. I-Penta-MAX}. This
corresponds to a solid angle of 0.01 of $4\pi$. A magnification of
$\times$20 is chosen. In opposite direction, a similar lens with
magnification of $\times$7 is used for single photon counting with
a photomultiplier\footnote{Electron-Tubes Inc., P25} (PMT). We
estimate an overall detection efficiency of 0.1\,$\%$ and
0.2\,$\%$ for the CCD and PMT, respectively. We typically obtain a
PMT count rate of $\sim$30\,kHz from a single ion, while the stray
light level is below 2\,kHz. The direction of the detection with
respect to the trap symmetry axis is $\{$-68$_{hor.}^\circ$,
0$_{vert.}^\circ\}$ and $\{$112$_{hor.}^\circ$,
0$_{vert.}^\circ\}$ for CCD and PMT, respectively.

\subsection{Laser setup for the qubit transition \label{LaserSetup729}}

Qubit operations are performed with laser light near 729~nm,
generated from a second Ti:Sapphire laser\footnote{Coherent Inc.,
899-21}. To obtain a high fidelity of gate operations this laser
source has to be stabilized in frequency and intensity to a high
degree. For frequency stability, we rely on a stable reference
cavity. Its length stability is guaranteed by a spacer from
ultra-low thermal expansion material (ULE) on which the cavity
mirrors (super-mirrors with a few ppm loss and transmission,
measured finesse of 2.4$\times$10$^5$) are optically contacted.
For additional stability, the cavity is suspended on wires in a
temperature stabilized UHV chamber. We derive a Pound-Drever-Hall
error signal\cite{DREVER83} and stabilize the laser frequency with
a servo bandwidth of $\leq$2.5~MHz obtaining a laser linewidth
$\leq100$~Hz \cite{ROHDE_DR}. The laser intensity is stabilized
using an AOM to about 1$\%(rms)$.

\begin{figure}[t]
\begin{center} \epsfig{file=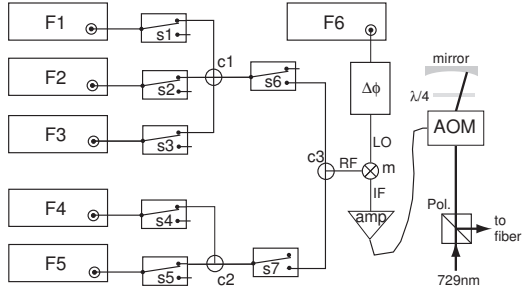,
width=0.8\linewidth} \caption{ \label{RFnetwork}
 \addtocounter{footnote}{3}
RF setup for control of 729\,nm laser: The output of
RF-synthesizers F1 to F5\protect\footnotemark\ is controlled by
switches ($s1$ to $s7$)\protect\footnotemark, added up with
combiners ($c1$ to $c3$)\protect\footnotemark\ and
mixed\protect\footnotemark\ with the output of the synthesizer
$F6$, which serves as local oscillator (LO). The phase of the LO
is controlled via a digital phase
shifter(DPS)\protect\footnotemark. The output (IF) is amplified
and fed to an acousto-optical modulator, operated in double-pass
configuration. The frequencies F1 to F6 are computer controlled
via GPIB. \addtocounter{footnote}{-5}
  \addtocounter{footnote}{-3} 
}
\end{center}
\end{figure}

The qubit operations require laser pulses with well defined phase,
frequency, intensity and duration. We modulate the output of the
Ti:Sapphire laser ($\sim$350~mW) with an AOM\footnote{Brimrose
Inc.,TEF-270-100} (see fig.~\ref{RFnetwork}) in double-pass
configuration. The radio frequencies and phases that are applied
to the AOM transfer directly to the light field\footnote{Due to
the double-pass configuration, the modulation of laser frequency
and phase is twice the applied RF modulation.}.

For maximum flexibility of the complex temporal pattern, we use
the scheme depicted in fig.~\ref{RFnetwork}.  Sideband ground
state cooling is performed with sources F4 and F5 at frequencies
resonant to the red sidebands $\omega_0$ and $\sqrt{3}\,\omega_0
$. The specific quantum gate sequence is composed of pulses on the
carrier and blue sideband of the bus mode, driven by the sources
F1 and F2, while F3 is used for the AC-Stark compensation (see
sect.~\ref{ACStark}). The computer digital output
card\footnote{J\"ager Inc., ADwin}, temporal resolution 1~$\mu$s,
32 channels, serves to switch the frequency sources and the
digital phase shifter. F6 compensates for the drift of the laser
reference cavity. The linear drift component of $\leq$ 10~Hz/s is
determined by comparison to the atomic resonance and
anticipatively corrected for. As we saturate the LO input of the
frequency mixer $m$, a small RF-transmission modulation of the
phase shifter for different $\Delta \phi$ does not convert into a
RF-intensity modulation at IF. In addition, this RF-setup allows
for the generation of multi-chromatic light fields, as necessary
e.g. for the compensation of the AC-Stark effect.
 \stepcounter{footnote}\footnotetext{Marconi Inc., Signal gen. 2023}
 \stepcounter{footnote}\footnotetext{Minicircuits Inc., ZYSW-2-50DR}
 \stepcounter{footnote}\footnotetext{Minicircuits Inc., ZSC-2-1}
 \stepcounter{footnote}\footnotetext{Minicircuits Inc., ZP-2}
 \stepcounter{footnote}\footnotetext{Lorch Inc., DP-1-8-370-5-77}

\subsection{Single ion addressing optics}\label{AddressingOptics}
For addressing individual ions, light at 729\,nm is spatially
filtered and transported by an optical fibre. A two-lens telescope
expands the 729~nm beam while an electro-optic
deflector(EOD)\footnote{LaserComponents Inc., ED2-730} in front of
the lenses controls the beam direction. We direct this expanded
laser beam (w$_0\sim$1cm) counterpropagating to the emerging
fluorescence towards the CCD, using a dichroic beam splitter and
focus it onto individual ions by using the same
lens\footnote{Nikon,~MNH-23150-ED-Plan-1.5x} as for imaging the
fluorescence light. The focused beam of up to 80~mW hits the ion
crystal under an angle of
$\{$68$_{hor.}^\circ$,$0_{vert.}^\circ\}$. The corresponding
single ion Lamb-Dicke factors are $\eta_{\rm{axial}}=0.033$ and
$\eta_{\rm{radial}}=0.040$, respectively. By varying the voltage
applied to the EOD we steer the focus at the ion position by more
than 10$\mu$m, large compared to the two-ion distance of
4.90$\mu$m\footnote{Projection of the two-ion distance of
5.29$\mu$m under 22$^\circ$} for
$\omega_\text{ax}/(2\pi)$=1.2~MHz. Additionally, the high-voltage
controller for the deflector can be preset to values which are
selected through digital input lines. The digital signals are
computer-generated by the same digital output board that controls
the RF pulses. Between different addressing positions, we
typically leave a settling time of 15\,$\mu$s. The determination
of the spatial resolution is discussed in sect.~\ref{Addressing}.

In order to provide a quantization axis and to split the Zeeman
components of the S$_{1/2}$ to D$_{5/2}$ transition, we compensate
the ambient magnetic field and generate a constant magnetic field
of 2.4~G under an angle of $\{$22$_{hor.}^\circ$,
0$_{vert.}^\circ\}$ which is thus perpendicular to the
$\overrightarrow{k}$-vector of the addressing light field. The
geometry and polarization of the light field at 729~nm allows the
excitation of $\Delta m = 0, \pm1$ and $\pm~2$
transitions\cite{ROOS_DR}.

\section{Preparative procedures and measurements}
\label{PROCEDURES}
 This section addresses the methods that are
used to prepare and manipulate the ions for a typical experimental
sequence. In a first step, the ions are initialized in a well
defined state using sympathetic sideband cooling and optical
pumping (sect.~\ref{GroundStateCooling}). Then, the ions are
individually manipulated on the qubit transition. During
manipulation on the sideband frequencies, the level-shifts due to
the AC-Stark effect need to be counteracted by additional laser
frequencies (sect.~\ref{ACStark}). Finally, the individual states
of the ions are detected by means of a CCD camera and a PMT
(sect.~\ref{QubitReadout}).


\subsection{Ground state cooling}
 \label{GroundStateCooling}
Each experimental cycle starts with the preparation of the ions in
a well defined initial state. The motional state of the two ion
crystal can be described by 6 different vibrational
modes\cite{JAMES98}. The axial and two radial center-of-mass modes
at $\omega_\text{ax}$ and $\omega_\text{rad}^{(x,y)}$ coincide
with the single-ion trap frequencies. The two radial rocking modes
and one axial breathing mode have frequencies of
$\omega_\text{R}^{(x,y)}= \sqrt{\omega_\text{rad}^{(x,y)}\,^2
-{\omega_\text{ax}}^2}$ and $\omega_\text{b}=\sqrt3
\,\omega_\text{ax}$, respectively. In our experiment, we have
chosen the breathing mode as the `bus-mode' for the quantum gate
and we therefore need to prepare it in the ground state
$\left|n_\text{b}=0\right>$. For the radial spectator modes,
Doppler cooling is sufficient as
$\eta_\text{rad}=0.04\ll$1\footnote{The final temperature is close
to the Doppler cooling limit if the UV light intensity is below
saturation}. However, the axial spectator mode\cite{WINELAND_B98}
at $\omega_\text{ax}$ is sideband cooled in addition to the
bus-mode.

The cooling cycle starts with a 2\,ms period of Doppler cooling on
the S$_{1/2}$ to P$_{1/2}$ transition at 397\,nm during which the
repumping laser on the D$_{3/2}$ to P$_{1/2}$ line (866\,nm) is
switched on.

After a short period of optical pumping into the
S$_{1/2}(m_J=-1/2)$ state (typically 30$\,\mu$s), the bus-mode and
the axial center of mass mode are sequentially sideband cooled
using the quadrupole transition at 729\,nm \cite{ROHDE01}. We
switch the 729\,nm-laser to one of the two ions and subsequently
perform a cooling cycle for 2\,ms and 6\,ms on the red sideband of
the center-of-mass mode and of the bus-mode, respectively. The
cooling rate is enhanced to several kHz by a quench laser on the
D$_{5/2}$ to P$_{3/2}$ transition at 854\,nm. During these
periods, the $\sigma^-$-beam is repeatedly pulsed on every 2\,ms
to recollect atoms that have been pumped to the
S$_{1/2}(m_J=+1/2)$ state.

With this procedure, we achieve a ground state population of the
bus-mode of about 99\%\cite{ROOS99,ROHDE01} and a coefficient of
$\eta^2\bar{n} \ll 1$ for all spectator modes. At the end of the
cooling cycle, a last optical pumping pulse initializes the ion
chain in the electronic ground state S$_{1/2}(m_J=-1/2)$.

\subsection{Addressing single ions}
\label{Addressing} As explained in section~\ref{AddressingOptics},
the focus position of the manipulation laser at 729\,nm is
controlled by the EOD.

\begin{figure}[t]
\begin{center}
\epsfig{file=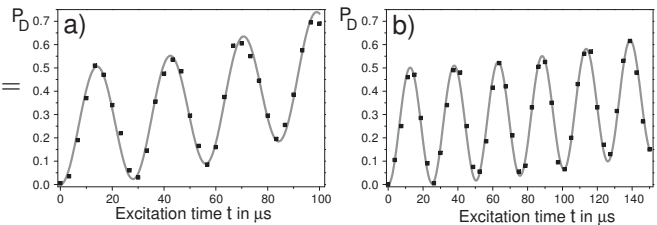,width=\linewidth}
\caption{\label{AddrCarrier2Ions} Rabi oscillations on the
carrier, performed on a two-ion crystal after ground state
sideband cooling of both the axial COM and the breathing mode. The
plots show the average excitation into the $|D\rangle$ state
(\,$({P_{D,1}+P_{D,2}})/2$) which is measured with the PMT. For
the data shown in (a), the laser is addressed onto the first of
the two ions, for (b) onto the second one. For the given
adjustment of the optics, the addressing error is different for
the two ions: (a) $\Omega_1=2\pi\cdot35.5(1)$~kHz,
$\Omega_2=2\pi\cdot2.46(7)$~kHz. We find an addressing error
$\Omega_1/\Omega_2=6.9(1)$\%, and a ratio of light intensities of
1:210. (b) $\Omega_1=2\pi\cdot39.7(2)$~kHz,
$\Omega_2=2\pi\cdot1.16(5)$~kHz, which corresponds to a ratio of
Rabi frequencies of $2.9(1)$\%, and 1:1200 for the light
intensities.}
\end{center}
\end{figure}

The quality of the addressing can be evaluated from Rabi
oscillations between the S$_{1/2}(m_J=-1/2)$ and the
D$_{5/2}(m_J=-1/2)$ state, that are driven on {\it one} out of two
ions. Residual laser light that reaches the second ion leads to a
slow Rabi oscillation of the second ion. From such measurement, we
infer the addressing error, i.e. the amount of unwanted qubit
rotation on the second ion which is present during a particular
one-qubit manipulation on the first ion, and vice versa.
Fig.~\ref{AddrCarrier2Ions} shows two typical excitation patterns
for such Rabi flops.

It is important to note that this addressing error does not
fundamentally limit the accuracy of one-qubit rotations. For the
current experiments, we have included this effect in the error
budget\cite{SCHMIDTKALER03}. It is, however, possible to
counteract the unwanted rotation on the second ion by an
additional laser pulse that is addressed to the second ion. The
remaining error on the first ion would then be of second order,
and even this contribution could be
eliminated by a clever choice of pulses. 
To make such counteraction possible, one has to determine the
phase difference between the laser light addressed directly to ion
1 and the residual light that generates the unwanted rotation on
ion 1 while the beam is addressed to ion 2.

\begin{figure}[t]
\begin{center}
\epsfig{file=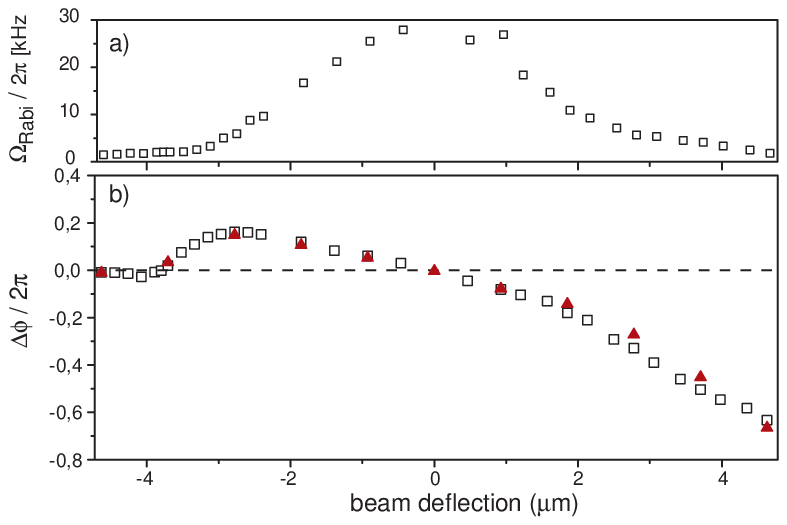,width=0.8\linewidth}
\caption{\label{fig:AddressingPhase} (a)  Rabi frequency of the
ion as a function of the deflector voltage $U_\text{defl}$.
\newline (b)Phase difference $\Delta\Phi$ between the addressed
($U_\text{defl}=0$) and the deflected beam, as perceived at the
ion's position. The different symbols represent independent
measurements. The data stem from a measuring period of more than 4
hours.}
\end{center}
\end{figure}

We have measured this phase difference with only one ion in the
trap. For this, we adjust the beam such that, without deflection,
the ion is centrally addressed. Figure~\ref{fig:AddressingPhase}a
shows the dependence of the Rabi frequency on the beam deflection.
The corresponding laser intensity is approximately given by a
Gaussian with a waist of 2.5\,$\mu$m\cite{GULDE_DR}. We now
perform a spin-echo experiment with the laser frequency tuned to
the carrier transition. The two framing pulses are performed with
a deflected beam and with the controlled phase set to 0 and $\pi$,
respectively. Because of the beam deflection, the ion feels the
laser phase $\Delta\Phi$ and $\Delta\Phi+\pi$. The center pulse is
directly addressed to the ion, with the controlled phase set to
$\phi$. If we define $R^x(\theta,\Phi)$\cite{GULDE03} to be the
qubit rotation by an angle $\theta$ about the horizontal axis
characterized by the polar angle $\Phi$, where $x$ denotes the
beam deflection, then the action on the atom can be described by:
 \begin{equation} \label{eq:spinecho}
 R_\text{echo}^{x} =
 R^x(\frac\pi2,\Delta\Phi)\,R^0(\pi,\Phi)\,R^x(\frac\pi2,\Delta\Phi+\pi)\mbox{.}
 \end{equation}
If the phase difference between the deflected and the addressed
beam $\Delta\Phi$ is equal to zero, then we expect no spin flip
for a phase $\Phi=\pm \pi/2$. Moreover, scanning the phase $\Phi$
yields an excitation from the $|S\rangle$ to the $|D\rangle$ state
of $P_D = \cos^2(\Phi-\Delta\Phi)$ which can be fitted to infer
the phase shift $\Delta\Phi$. The dependence of this phase shift
on the beam deflection is shown in
fig.~\ref{fig:AddressingPhase}b.


We attribute the linear part of the phase shift to the elongation
of the optical path within the EOD. Such behavior is expected if
the laser beam is not ideally aligned with the EOD axis. For beam
deflections larger than $\pm$2\,$\mu$m, $\Delta\Phi$ depends no
longer in a linear way on the deflection. We suppose that this is
due to light which does not travel through the optical system
along the ideal path and therefore has a phase different from the
Gaussian part of the beam. Such a hypothesis is supported by the
small pedestal below the Gaussian profile in
fig.~\ref{fig:AddressingPhase}.

As an important result we note that the phase difference between
the deflected and the addressed beam is well defined and stable
over long periods of time, even for large deflections.
This offers the possibility to reduce the effect of the addressing
error in future experiments.

\subsection{AC Stark compensation}
 \label{ACStark}
As the ions which represent the qubits are not ideal two-level
systems, the manipulation of the qubit states can be perturbed by
non-resonant coupling to other levels. In particular, for
manipulations on the vibrational sideband, the coupling to the
carrier\cite{STEANE00} is so strong that it induces important
light shifts (AC Stark shifts) on the qubit levels. As this would
perturb their phases, the light shift needs to be compensated by
an additional laser frequency of appropriate power and
detuning\cite{HAFFNER03}. The compensating light field is
generated by a frequency F3, also applied to the double-pass AOM
in the 729\,nm beam (see section \ref{LaserSetup729}). Using the
same laser beam as a source, this setup ensures that laser power
fluctuations or changes in the beam alignment are not converted
into phase fluctuations.

\subsection{Phase gate and composite laser pulses\label{phasegate}}

The central quantum-logic operation in the Cirac-Zoller CNOT-gate
is a one-ion phase gate where the sign of the electronic qubit is
switched conditional on the vibrational state. In the
computational subspace ($\ket{D,0}$, $\ket{D,1}$,$\ket{S,0}$,
$\ket{S,1}$), this gate is described by a diagonal matrix with the
entries (1,-1,-1,-1)\footnote{This transformation is the standard
phase gate up to an overall phase factor of $-1$.}.

\begin{figure}[t]
\begin{center}
\epsfig{file=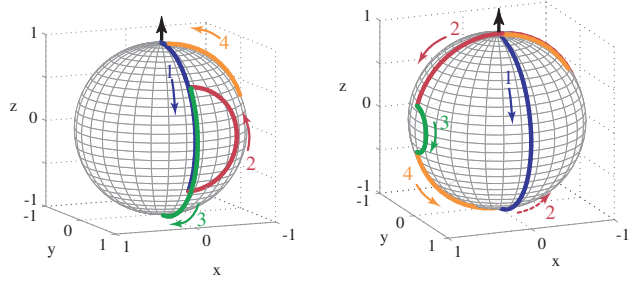,width=0.99\linewidth}
 \caption{\label{PHASE}
Bloch sphere trajectories for the composite phase gate,
$R_\text{phase}$. Left: Bloch sphere for the
quasi-two-level-system $|S,0\rangle \leftrightarrow |D,1\rangle$.
The initial state is $|S,0\rangle$, indicated by the black arrow.
Pulse $R_1$ of the
sequence 
rotates the state vector about the $x$-axis by $\pi/\sqrt{2}$.
$R_2$ accomplishes a $\pi$-rotation about the $y$-axis. It
therefore transforms the state to its mirror image about the
$x$-$y$-plane. Consequently, $R_3$, which is identical to $R_1$,
rotates the state vector all the way down to the bottom of the
sphere. $R_4$, just like $R_2$, represents a $\pi$-rotation about
the $y$-axis. The final state identical to the initial one, except
the acquired phase factor $-1$. \newline Right: The same laser
pulse sequence acting in the $|S,1\rangle \leftrightarrow
|D,2\rangle$ subspace. Again, the final state identical to the
initial one, except the acquired phase factor $-1$. }
\end{center}
\end{figure}

Excitation on the blue motional sideband leads to a pairwise
coupling between levels $\ket{S,n}\leftrightarrow\ket{D,n+1}$
except for the level $\ket{D,0}$. For the phase gate we perform an
effective  $2\pi$-pulse on the two two-level systems
($\ket{S,0}\leftrightarrow\ket{D,1})$ and
$(\ket{S,1}\leftrightarrow\ket{D,2})$ which changes the sign of
all computational basis states except for $\ket{D,0}$). Since the
Rabi frequency depends on $n$, we need to use a composite-pulse
sequence\cite{CHILDS00} instead of a single blue sideband pulse.
The sequence is composed of four sideband pulses
$R_4\,R_3\,R_2\,R_1$ and can be described by
\begin{eqnarray}
  R_\text{phase} &=&
  R^+(\pi\sqrt{n+1},0)\,R^+(\pi\sqrt{\textstyle\frac{n+1}{2}},\pi/2)\,\nonumber
  \\
&& \cdot
R^+(\pi\sqrt{n+1},0)\,R^+(\pi\sqrt{\textstyle\frac{n+1}{2}},\pi/2)
\end{eqnarray}
where $n$ denotes the lower vibrational quantum number of the two
coupled states and $R^+$\cite{GULDE03}, similar to $R$ in equation
\ref{eq:spinecho}, denotes a rotation induced by coupling to the
upper motional sideband. Figure~\ref{PHASE} illustrates the
evolution of the Bloch vectors during the phase gate and provides
a step-by-step picture of the process\footnote{The Bloch-sphere
picture doesn't give complete information on the phases picked up
during the evolution. Those have to be computed using a matrix
representation.}.

 It may be helpful to interpret this evolution in terms of spin-echos.
For the system $(\ket{S,0}\leftrightarrow\ket{D,1})$, the first
three pulses constitute a spin-echo experiment where the
$\pi$-pulse in the middle assures that the overall evolution is
the one of a $\pi$-pulse, despite the rotation angle of
$\pi/{\sqrt2}$. This evolution is followed by an additional $\pi$
pulse which completes the $2\pi$-rotation. For the second
two-level system $(\ket{S,1}\leftrightarrow\ket{D,2})$, the
sequence starts with a $\pi$-pulse that is followed by the
spin-echo-type $\pi$-rotation.

The phase gate is transformed into a CNOT operation if it is
sandwiched in between two $\pi/2$ carrier pulses,
$R_{\rm{CNOT}}=R(\pi/2,0)\,R_{\rm{phase}} \, R(\pi/2,\pi)$.

\subsection{Qubit readout}
\label{QubitReadout}

For detection of the internal quantum states, we excite the
S$_{1/2}$ to P$_{1/2}$ dipole transition near 397\,nm and monitor
the fluorescence. This measurement collapses the wave function
onto the two states $|S\rangle$ and $|D\rangle$, fluorescence
indicating the S$_{1/2}$ state, no fluorescence revealing the
D$_{5/2}$ state. By repeating the experimental cycle 100 times we
find the average state populations.

\begin{figure}[ht]
\epsfig{file=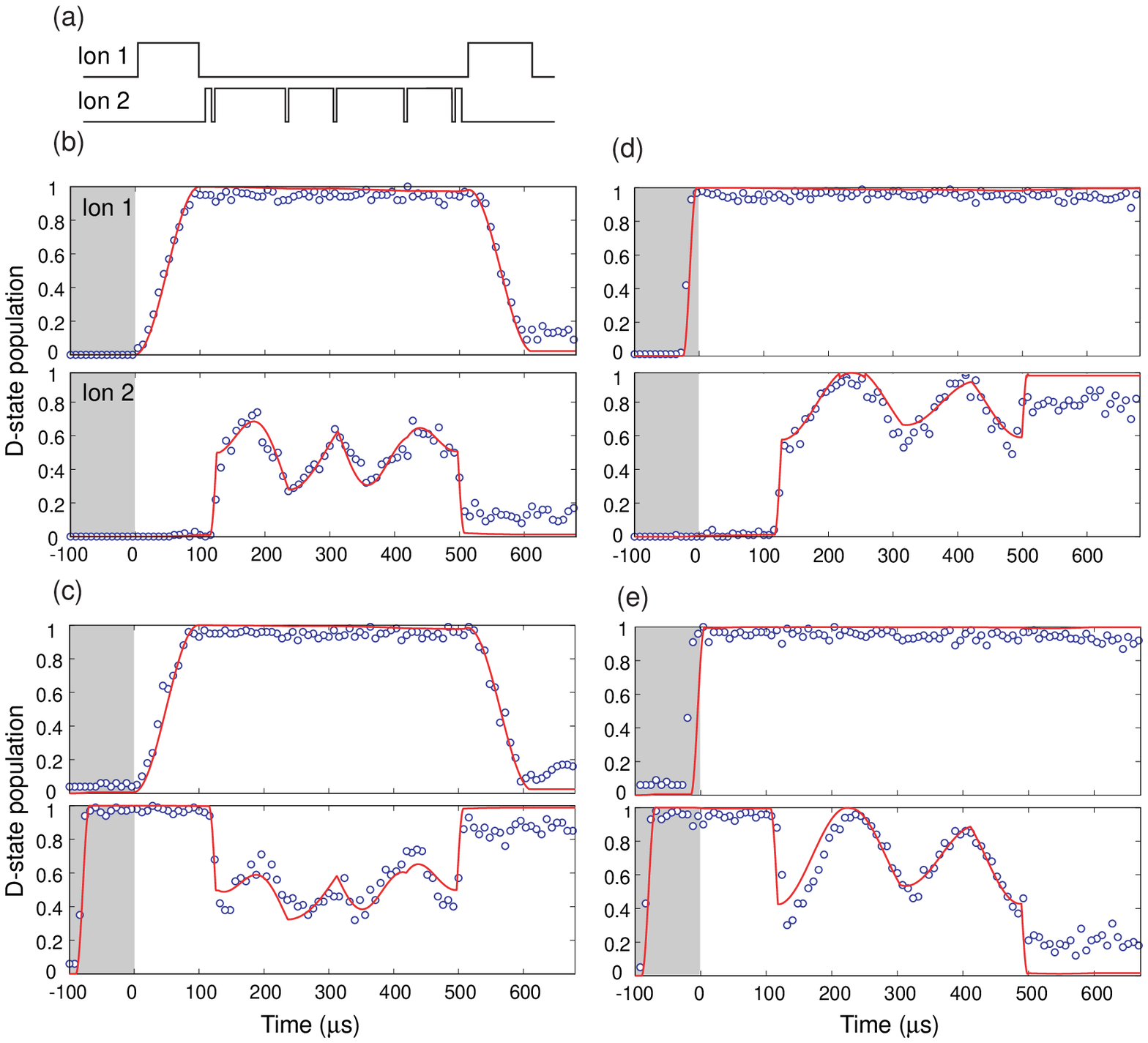, width=0.99\linewidth} \caption{State
evolution of both qubits
$|control,target\rangle=|ion\,1,ion\,2\rangle$ under the CNOT
operation. First, we initialize the quantum register in the state
(b) $|S,S\rangle$, (c) $|S,D\rangle$, (d) $|D,S\rangle$, or (e)
$|D,D\rangle$ (shaded area, $t\le0$). Then, the quantum gate pulse
sequence (a) is applied: After mapping the first ion's state
(control qubit) with a $\pi$-pulse of length 95\,$\mu$s to the
bus-mode, the single-ion CNOT sequence (consisting of 6
concatenated pulses) is applied to the second ion (target qubit)
for a total time of 380\,$\mu$s. Finally, the control qubit and
bus mode are reset to their initial values with the concluding
$\pi$-pulse applied to the first ion. To follow the temporal
evolution of both qubits during the gate, the pulse sequence (a)
is truncated and the $|D\rangle$ state probability is measured as
a function of time. The solid lines indicate the theoretically
expected behavior. Input parameters for its calculation are the
independently measured Rabi frequencies on the carrier and
sideband transitions and the addressing error.\label{BasisCNOT}}
\end{figure}

By means of an intensified CCD camera, the fluorescence can be
monitored separately for each ion. Typical exposure times range
from 23\,ms (data of fig.~\ref{BasisCNOT}) down to 10\,ms (all
other data). The fluorescence is integrated over an area of
~3\,$\mu$m\,x\,3\,$\mu$m around the ions' center, which
corresponds to ~3\,x\,3 pixels on the CCD camera. With this
method, state detection of each qubit is performed with an
accuracy of ~0.98, the residual error of ~2$\%$ resulting from
spurious fluorescence light of the adjacent ion (cross-talk) or,
in the case of the 23\,ms collection time, from spontaneous decay.

If no information on a particular qubit is needed (as in the
experiments on the addressing error), we use the signal of a
photomultiplier tube to infer the overall state population. In
this case, we reduce the exposure time to 3.5\,ms.

\section{Two-ion universal gate \label{TWOIONGATE}}
For the two-qubit CNOT operation, Cirac and Zoller proposed to use
the common vibration of an ion string to convey the information
for a conditional operation (bus-mode)\cite{CIRAC95}. Accordingly,
the gate operation can be achieved with a sequence of three steps
after the ion string has been prepared in the ground state
$|n_b=0\rangle$ of the bus-mode. First, the quantum information of
the control ion is mapped onto this vibrational mode, the entire
string of ions is moving and thus the target ion participates in
the common motion. Second, and conditional upon the motional
state, the target ion's qubit is inverted\cite{MONROE95}. Finally,
the state of the bus-mode is mapped back onto the control ion.
Note, that this gate operation is not restricted to a two-ion
crystal since the vibrational bus mode can be used to interconnect
any of the ions in a large crystal, independent of their position.

We realize this gate operation\cite{SCHMIDTKALER03} with a
sequence of laser pulses. A blue sideband $\pi$-pulse,
$R^+(\pi,0)$, on the control ion transfers its quantum state to
the bus-mode. Next, we apply the CNOT operation $R_{\rm{CNOT}}$ to
the target ion, see sect.~\ref{phasegate}. Finally, the bus-mode
and the control ion are reset to their initial states by another
$\pi$-pulse $R^+(\pi,\pi)$ on the blue sideband. We apply the gate
to all computational basis states and follow their temporal
evolution, see fig.~\ref{BasisCNOT}. The desired output is reached
with a fidelity of 71 to 77\%.

\begin{figure}[ht]
\epsfig{file=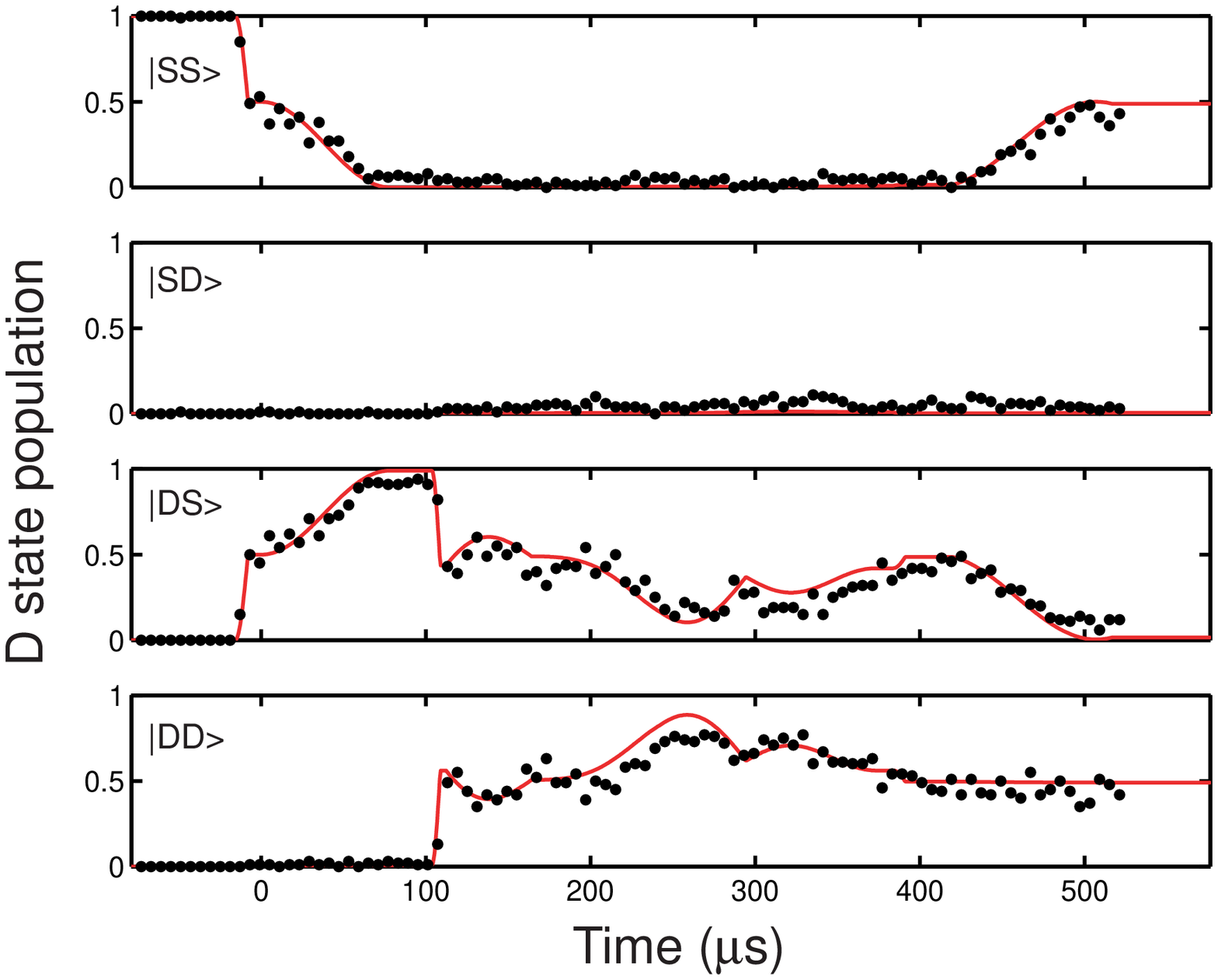, width=0.99\linewidth}
\caption{\label{bellfast}Left: the controlled-NOT gate operation
$R_{\rm{CNOT}}$ is performed with ions initially prepared in
$|S+D,S\rangle$. The data points represent the probability for the
ion string to be in the state indicated on the right-hand side by
the corresponding CCD image, during the execution of the gate. The
measurement procedure is the same as in fig.~\ref{BasisCNOT}.
Right: CCD images of the fluorescence of the two-ion crystal as
measured in different logic basis states:
$|SS\rangle,|SD\rangle,|DS\rangle,$ and $|DD\rangle$. The ion
distance is 5.3~$\mu$m.}
\end{figure}

\begin{figure}[ht]
\centerline{\epsfxsize=2.0in\epsfbox{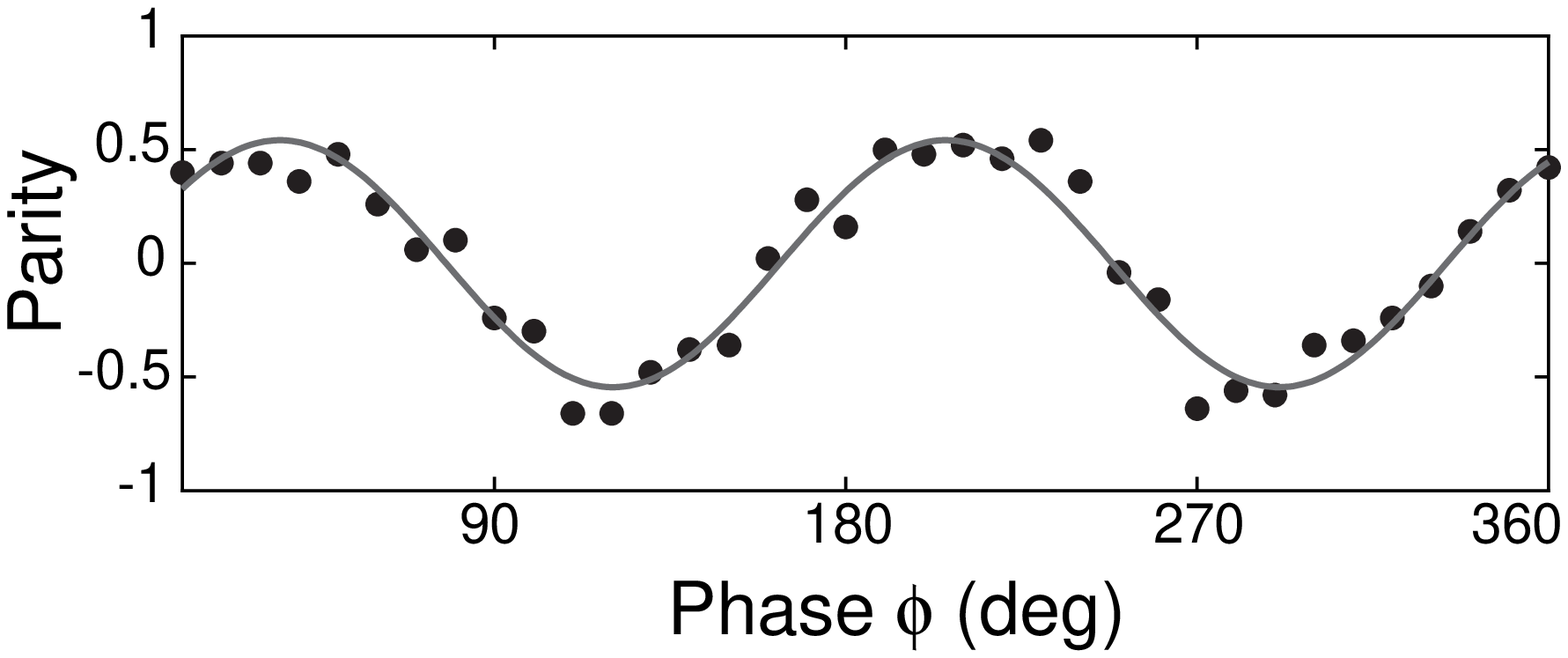}}
\caption{\label{parity} Analysis of the entangled output state of
a  CNOT: After the gate operation, we apply $\pi$/2-pulses on the
carrier transition, with a phase $\phi$, to both ions, and measure
the parity $P = P_{SS} + P_{DD} - (P_{SD} + P_{DS})$ as a function
of the phase. The quantum nature of the gate operation is proved
by observing oscillations with $\cos{2\phi}$, whereas a
non-entangled state would yield a variation with $\cos{\phi}$
only. The observed visibility is 0.54(3).}
\end{figure}

If the qubits are initialized in the superposition state
$|$control, target$\rangle= |S+D,S\rangle$, the CNOT operation
generates an entangled state $|S,S \rangle + |D,D\rangle$. The
corresponding data are plotted in fig.~\ref{bellfast}, left side.
At the end of the sequence, near t=500\,$\mu$s, only the states
$|S,S\rangle$ and $|D,D\rangle$ are observed with $P_{SS}$=0.42(3)
and $P_{DD}$=0.45(3). The phase coherence of both these components
is verified by applying additional analysis $\pi$/2-pulses on the
carrier transition followed by the projective measurement. From
the observed populations prior to the analyzing pulses and the
contrast of the oscillation, see fig.~\ref{parity}, we calculate
the fidelity according to the prescription given in Sackett et
al.\cite{SACKETT00}, and find a gate fidelity of
0.71(3)\cite{SCHMIDTKALER03}.

\subsection{Limitations}
The gate fidelity is well understood in terms of a collection of
experimental imperfections\cite{SCHMIDTKALER03}. Most important is
dephasing due to laser frequency noise and due to ambient magnetic
field fluctuations that cause a Zeeman shift of the qubit
levels\cite{SCHMIDTKALER03b}. As quantum computing might be
understood as a multi-particle  Ramsey interference experiment, a
faster execution of the gate operation would help to overcome this
type of dephasing errors. However, a different type of error
increases with the gate speed: With higher Rabi frequencies, the
off-resonant excitation of the nearby and strong carrier
transition is increasingly important\cite{STEANE00} even if the
corresponding phase shift is compensated. Additional but minor
errors are due the addressing imperfection, residual thermal
excitation of the bus mode and spectator modes and laser intensity
fluctuations. In future, higher trap frequencies and the use of
hyperfine ground states coupled by Raman transitions for the qubit
will improve the gate fidelity.

\section{Summary and outlook}
 \label{Summary}
We have demonstrated universal single and two-qubit operations on
an elementary quantum processor. The results shows the feasibility
of ion trap technologies for QC.  Recently, tomographic quantum
state reconstruction has been implemented\cite{ROOS03}. For
process -- or gate -- tomography, an appropriate set of initial
states and their superpositions may be processed and the output
tomographically be analyzed. The application of this technique
will allow to fully characterize any kind of quantum evolution and
deduce the underlying Hamilton operator. This will help to devise
more accurate and more complex quantum operations.

This work is supported by the Austrian `Fonds zur F\"orderung der
wissenschaftlichen Forschung', by the European Commission, and by
the `Institut f\"ur Quanteninformation GmbH'.


\end{document}